\documentclass[pre,letterpaper,twocolumn,floatfix,superscriptaddress]{revtex4}
\usepackage{graphicx}

\begin{document}

\title{Improved Local Lattice Approach for Coulombic Simulations}

\author{A.~Duncan}
\affiliation{Department of Physics and Astronomy\\
University of Pittsburgh, Pittsburgh, PA 15260}

\author{R.D.~Sedgewick}
\author{R.D.~Coalson}
\affiliation{Department of Chemistry\\
University of Pittsburgh, Pittsburgh, PA 15260}

\begin{abstract}
An improved approach to the simulation of strongly fluctuating Coulomb
gases, based on a local lattice technique introduced by Maggs and
Rossetto \cite{maggs:prl}, is described and then tested in a problem
of biophysical interest.  The low acceptance rates for charged
particle moves in regimes of physical interest are increased to a
serviceable level by use of a coupled particle-field update procedure
in the new method. Sensitivity of the results to lattice
discretization effects is also studied using asymmetric lattices.

\end{abstract}
\maketitle

\section{Introduction}

The simulation of systems containing a large number of mobile charged
entities, in which long range electrostatic forces play a central
dynamical role, is of critical importance in modern chemical physics
and biophysics.  In many cases, the computational load is dominated by
the evaluation of the electrostatic energy of the system, where the
long-range character of the Coulomb interaction greatly complicates the
development of efficient algorithms that scale with system size in a
way that permits study of systems of biophysical
interest. Developments in supercomputing technology aimed at
large-scale biophysical simulations, such as the IBM BlueGene project
\cite{bluegene}, where massively parallel assemblies of processor
nodes are coupled via a three-dimensional toroidal topology, suggest
that algorithms based on a local energy functional will be much more
efficiently executed on the next generation of high-end computing
platforms than those involving long-range nonlocal effects.

  Recently, Maggs and collaborators \cite{maggs:prl} have suggested an
ingenious procedure for removing the nonlocal (long-ranged) Coulomb
term in equilibrium simulations of Coulomb gases. By using a
completely local Hamiltonian for a system of mobile charged particles
interacting with the electrostatic field, one avoids the unpleasant
scaling characteristics of conventional Coulomb gas
simulations. Unfortunately (as pointed out by these authors themselves
\cite{maggs:acc_rate}), the algorithm they propose runs into serious
acceptance problems in regions of physical interest (basically, for
strongly fluctuating systems).  In this paper, we study the origin of
these acceptance problems and propose an improved algorithm that
allows useful simulations of strongly fluctuating systems in which
mean-field (or Poisson-Boltzmann) methods break down.

  In Section 2 we briefly review the original technique of Maggs et
al, and explain the origin of the acceptance difficulty for charged
particle moves.  In Section 3 we explain the modified update procedure
designed to cure, or at least ameliorate, the acceptance
problem for particle moves. In brief, the crucial point is to
implement a coupled particle-field update in which the electrostatic
field is allowed to readjust itself in tandem with charged particle
moves in response to the changed electrostatic environment.  In
Section 4, the improved procedure is tested in detail on a system that
has been extensively studied in the literature
\cite{podgornik88,stevens90,guldbrand84,moreira02}: the osmotic
pressure of charged plates (or membranes) separated by an ionic
fluid. Finally, in Section 5 we briefly summarize our conclusions.

\section{Local Lattice Hamiltonians for Coulomb Gas Problems}            
 
The difficulties incurred by the nonlocal nature of the Coulomb
interaction in realistic simulations of large systems (for example,
for large biomolecular systems) are well known: the computational cost
increases as the square of the number of charged constituents, and
although various techniques (Ewald summation, fast Fourier transforms
etc. \cite{Leach}) can be employed to improve this scaling, the
resulting complications in the algorithm often mean that the
computation of the electrostatic energy still consumes essentially all
of the computational effort, greatly limiting the size of the systems
and (in the case of molecular dynamics simulations) the time frames
over which the simulations can be extended. These techniques also have
difficulties modeling a non-uniform dielectric constant, which is an
important feature of many biophysical
systems\cite{Bashford,Zacharias,Spassov} as the dielectric constant in
proteins is $\sim 2-8$ while the dielectric constant of water is
$\sim80$.  In the case of systems at equilibrium, it has been known
for some time \cite{podgornik88,Edwards59,Edwards62,tonyrob} that the
nonlocal Coulomb interaction can be replaced by a completely local
interaction via a Hubbard-Stratonovich transformation, yielding a path
integral formalism that connects naturally with the Poisson-Boltzmann
mean-field theory.  Unfortunately, for strongly fluctuating systems
perturbation theory (saddle-point expansions) breaks down in this
approach, and a direct numerical simulation is obstructed by a severe
sign problem.
   
Recently, Maggs and collaborators \cite{maggs:prl} have proposed an
alternative, purely local approach to the simulation of charged
condensed systems. They exploit the fact that the nonlocality of the
Coulomb interaction is a consequence of a particular choice of gauge
for describing the electromagnetic field, whereas the physically
relevant quantity --the electrostatic energy of the system-- must
clearly be a gauge-invariant object. They propose that the
electromagnetic field be simulated in terms of gauge-invariant objects
(specifically, the electric field), represented on a discrete spatial
lattice. In this respect, the method proposed is essentially the same
as that employed for over 20 years by elementary particle theorists
attacking the problem of strong interactions with the technique of
lattice quantum chromodynamics. For a review see Ref.~\onlinecite{LQCD}. The main distinction here
is that the gauge theory involved is the abelian one of Maxwellian
electrodynamics, magnetic effects are not relevant, and the
formulation used is a noncompact one (i.e. the electric field
variables take unbounded values).

Let us briefly recall the salient points of the formalism of Maggs et
al \cite{maggs:prl}. The canonical partition function for a set of 
mobile charges
${e_{i}}$ at locations $\vec{r}_{i}$ at inverse temperature $\beta$ in
a medium of dielectric constant $\epsilon$ may be written
\begin{equation}
   \label{eq:PartFunc}
   Z = \int\prod_{i=1}^{N}d\vec{r}_{i}\,{\cal D}\vec{E}(\vec{r})\prod_{\vec{r}}\delta\left(\vec{\nabla}\cdot\vec{E}-
   \frac{4 \pi}{\epsilon}\rho(\vec{r})\right)e^{-\frac{\beta\epsilon}{8\pi}\int d\vec{r}\vec{E}(\vec{r})^{2}}
\end{equation}
where the charge density $\rho(\vec{r})$  is shorthand for
\begin{equation}
     \rho(\vec{r}) \equiv \sum_{i}e_{i}\delta(\vec{r}-\vec{r}_{i})
\end{equation}
The delta function constraint in Eq.~\ref{eq:PartFunc} enforces Gauss'
Law, so that the electric fields integrated over correspond to the
particle locations specified through the density function $\rho$.  The
formulation is manifestly local, as both the energy functional and
Gauss' Law constraint are so. There is no requirement that the
electric fields integrated over be irrotational, and in fact they are
not; as shown by Maggs et al.\cite{maggs:prl}, the transverse part of
the electric field simply decouples from the particle sector and
contributes an irrelevant overall prefactor to $Z$.  Because this
formalism is local, it is easily extended to model physical systems
with a non-uniform dielectric constant \cite{maggs:dielectric}.
    
The functional integral over electric field in Eq.~\ref{eq:PartFunc}
can be given a precise definition by introducing a spatial cubical
lattice, which we shall for the time being take to be a grid of
$L^{3}$ points, with lattice spacing $a$ (in all directions: the
modifications needed in case of an asymmetric lattice are discussed
below) and periodic wrap-around boundary conditions in all three
spatial directions. The charges $e_{i}, i=1,..N$ are assumed to be
integer multiples of a basic unit of charge, $e_{i}=z_{i}e$, $z_{i}$
integer, and reside on the sites of the lattice. The component
$E_{\mu}(\vec{n})$ of electric field in direction $\mu$ at lattice
site $\vec{n}$ is associated with a real-valued field $E_{l}$ on the
oriented link $l$ from $\vec{n}$ to $\vec{n}+\hat{\mu}$. Discretizing
the 3-dimensional integral for the electrostatic energy in the obvious
way, we find
\begin{equation}
     \label{eq:discH}
      H\equiv\frac{\epsilon}{8\pi}\int d\vec{r}\vec{E}(\vec{r})^{2}\rightarrow
      \frac{a^{3}\epsilon}{8\pi}\sum_{l}E_{l}^{2}
\end{equation}

 The implementation of the simulation is simplified by introducing dimensionless
variables to the greatest extent possible, so we define $\hat{E_{l}}\equiv\frac{\epsilon a^{2}}{4\pi e}E_{l}$ and a rescaled inverse temperature $\hat{\beta}\equiv \frac{4\pi e^{2}\beta}{\epsilon a}$,
in terms of which the energy becomes
\begin{equation}
\label{eq:Hrescale}
    H = \frac{\hat{\beta}}{2}\sum_{l}\hat{E}_{l}^{2}
\end{equation}
while the Gauss' Law constraint takes the simple form
\begin{equation}
\label{eq:Gauss}
       \sum_{l}\hat{E}_{l} = z_{i}
\end{equation}
for the sum of outgoing link fields  from any site containing a charged particle of
charge $z_{i}e$.  

 The simulation of the system defined by the energy function in Eq.~\ref{eq:Hrescale} and 
the constraint in Eq.~\ref{eq:Gauss} can in principle be accomplished by the following
algorithm:
\begin{enumerate}
\item  Pick starting lattice locations (possibly randomly) for the $N$ particles of charge $z_{i},i=1,..N$.
 Then solve  Gauss' Law for these fixed charge locations to obtain a starting
configuration of electric link field variables satisfying the Gauss constraint.  This can
easily be done by standard numerical relaxation methods \cite{NR}.
\item Update the electric fields by shifting all link variables along a complete set of
independent closed paths by constant shifts, using either Metropolis or heat bath
procedures to accept/reject proposed shifts. The simplest version of this is simply
to consider all plaquettes (unit squares) on the lattice, shifting the 4 link fields ordered
around the plaquette by the same random amount $\alpha$, the range of $\alpha$ set so that there is a reasonable acceptance rate for the move.
Such a shift clearly maintains the Gauss' Law constraint.
\item Update particle locations by visiting in turn every site $ \vec{n}$ containing a charged particle
of charge $z_{i}$.
A particle move to the neighboring site $\vec{n}+\hat{\mu}$ in a random direction $\mu$ is then considered,
where the particle move is accompanied with a shift of the electric field $E_{l}$ on the link
$l=(\vec{n}\rightarrow\vec{n}+\hat{\mu})$
\begin{equation}
\label{eq:particlemove}
   \hat{E}_{l} \rightarrow \hat{E}_{l}-z_{i}
\end{equation}
in order to maintain the constraint in Eq.~\ref{eq:Gauss}. Here also one can employ either Metropolis or
heat-bath accept/reject procedures.
\end{enumerate}
%

  The inclusion of additional force fields, for example soft or hard
exclusion potentials modeling a finite size for the particles, is, in
principle, completely straightforward in this framework.  When
particles are packed closely together, or the potential changes
rapidly over the scale of a lattice spacing, then it is important to
verify that the observed phenomena are not distorted by lattice
discretization effects.  It is useful to be able to study the effects of
lattice discretization in such situations by introducing asymmetric
lattices, in which the lattice spacing in the various directions
differs. As a specific example, consider a situation in which we may
desire a finer discretization in the x-direction, relative to the y
and z directions, $a_{x}<a_{y}=a_{z}\equiv a$. One readily verifies
that with the choice of the dimensionless variables
\begin{equation}
\label{eq:asymmex}
  \hat{E}_{l} \equiv \left\{ \begin{array}{ll}
     \frac{\epsilon a^{2}}{4\pi e}E_{l} & l \in L_x \\
     \frac{\epsilon a a_{x}}{4\pi e}E_{l} & l \in L_y \cup L_z
\end{array} \right.
\end{equation}
where $L_\alpha$ is the set of links in the $\alpha$ direction and, as before,
\begin{equation}
  \hat{\beta} = \frac{4 \pi e^{2}}{\epsilon a}\beta
\end{equation}
 the energy function becomes
\begin{equation}
\label{eq:Hfuncasymm}
  H = \frac{\hat{\beta}}{2}\left(\sum_{l \in L_x}\frac{a_{x}}{a}\hat{E}^{2}_{l}
  +\sum_{l \in L_y}\frac{a}{a_{x}}\hat{E}^{2}_{l}+\sum_{l \in L_z}\frac{a}{a_{x}}\hat{E}^{2}_{l}\right)
\end{equation}
while the Gauss' Law constraint retains its original form given in Eq.~\ref{eq:Gauss}.

  Unfortunately, despite the appealing simplicity of the simulation
procedure outlined above, in physically realistic situations involving
strongly charged systems the method proves impractical, for reasons
we now explain. The dimensionless inverse temperature variable
$\hat{\beta}$ is typically large compared to unity (in the charged
plate/membrane problem considered in Section 4, the value is 87.1), so
that typical values for the electric field link variables are small
compared to unity. On the other hand, executing a particle move across
a link via Eq.~\ref{eq:particlemove} shifts the electric field
variable on that link by an integer, and this generally leads to an
unacceptable energy cost (on the order of $\hat{\beta}$).  In the
univalent case ($z_{i}=\pm 1$), acceptance rates for particle moves
are of the order of 10$^{-4}$, while for divalent ions ($z_{i}=\pm 2$)
the acceptance rate is at best of order 10$^{-6}$.  Thus the
unmodified procedure of Maggs et al.\  is clearly not a practical
approach in situations approximating real biophysical systems.  In the
next section, we discuss a modified simulation algorithm in which
this problem is ameliorated to an acceptable level.

\section{Solving the Particle Move Problem: A Coupled Update Procedure }                             
\label{sec:updateproc}

  The problem of very inefficient particle moves mentioned in the
preceding section needs to be resolved before the local Hamiltonian
method can be applied fruitfully to realistic problems with strongly
fluctuating Coulomb gases. Recall that the Hamiltonian, as a function of
the electric field variables $E_{l}$ defined on the links $l$ of the
lattice takes the form
\begin{equation}
\label{eq:HE}
   {\cal H} = \frac{\hat{\beta}}{2} \sum_{l} \hat{E}_{l}^{2}
\end{equation}
where the dimensionless inverse-temperature variable, $\hat{\beta}$,
is quite large for the systems that we are interested in studying (in
the range of 50-100). As discussed above, the vast majority of
particle moves with such a Hamiltonian have a high energy cost leading
to an unacceptably low acceptance rate.  In this section we will show
that this problem can be substantially ameliorated-- though not
completely eliminated-- by a coupled update procedure in which
electric field values on all the plaquettes containing the link along
which the particle move is attempted are simultaneously adjusted to
reflect the changed electrical environment resulting from the particle
move.

\begin{figure}
\centerline{\includegraphics[width=3.5in]{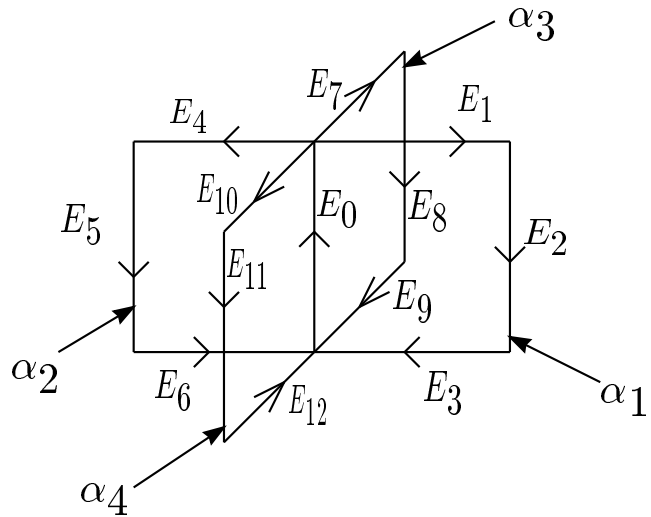}}
\caption{Local field environment for coupled particle move updates} 
\label{fig:plaquettes}
\end{figure}   

  As an example of a simple procedure that can considerably improve the
acceptance rate for particle moves, consider the situation illustrated
in Fig.~\ref{fig:plaquettes}.
Here we are considering the move of a unit charge particle from the
beginning (bottom) site to the end (top) site of the central link
associated with field variable $E_0$. In conjunction with the particle
move, we consider simultaneous electric field updates corresponding to
plaquette variable shifts $\alpha_1,\alpha_2, \alpha_3,\alpha_4$ on
the four plaquettes containing the link $E_0$, as indicated in the
figure. Such a combined move changes the energy associated with the
illustrated region from
\begin{equation}
\label{eq:hbefore}
   {\cal H}_{\bf before} = \frac{\hat{\beta}}{2} \sum_{l=0}^{12} \hat{E}_{l}^{2}
\end{equation}
to
\begin{eqnarray}
\label{eq:hafter}
   {\cal H}_{\bf after} &=& \frac{\hat{\beta}}{2} \Bigg\{ (\hat{E}_0+\sum_{i=1}^{4}\alpha_{i}-1)^{2}\\
 &+&\sum_{l=1}^{3}(\hat{E}_{l}+\alpha_1)^{2}
+\sum_{l=4}^{6}(\hat{E}_{l}+\alpha_2)^{2}\nonumber\\
&+&\sum_{l=7}^{9}(\hat{E}_{l}+\alpha_3)^{2}
+\sum_{l=10}^{12}(\hat{E}_{l}+\alpha_4)^{2}\Bigg\}\nonumber
\end{eqnarray}
  In practice one finds that the electric field variables equilibrate
to values which are small compared to unity: in the approximation
where we simply set $\hat{E}_{l}=0$ in Equations \ref{eq:hbefore} and
\ref{eq:hafter}, the energy cost of the combined move becomes
\begin{equation}
\label{eq:zeroE}
  \Delta H = \frac{\hat{\beta}}{2}\left((\sum_{i=1}^{4}\alpha_{i}-1)^{2}+3(\alpha_{1}^{2}+\alpha_{2}^{2}+
\alpha_{3}^{2}+\alpha_{4}^{2})\right)
\end{equation}
Minimizing Eq.~\ref{eq:zeroE} with respect to the $\alpha_i$, we find that the choice $\alpha_{i}=1/7$
gives the minimum energy cost 
\begin{equation}
  \Delta H_{\bf min} = \frac{3}{7}\, \frac{\hat{\beta}}{2} \approx 0.43\, \frac{\hat{\beta}}{2}
\end{equation}
as opposed to the cost $\hat{\beta}/2$ if the particle move is
unaccompanied by any readjustment of nearby link fields. As we shall
see in the next section, if $\hat{\beta}$ is large, this is enough to
 increase the acceptance rate for particle moves to a level
where configurations can be decorrelated at an acceptable rate. Thus,
a quick and easily implementable improvement of the basic algorithm
can be obtained by a Metropolis accept/reject step in which the
choices for a particle move on a chosen link are (a) do nothing (to
particle or fields), or (b) perform the combined update in which the
particle is transferred to the end site of the link and the fields
around the four intersecting plaquettes are shifted by $1/7$. It is
clear that the energy cost can be further reduced by allowing
readjustments of plaquettes adjacent to those depicted in
Fig.~\ref{fig:plaquettes}.  In particular, if the link $E_0$ in
Fig.~\ref{fig:plaquettes} corresponds to the $z$ direction, then
adding the remaining $xz$ and $yz$ plaquettes that contain the links
$E_2$, $E_5$, $E_8$ and $E_{11}$, and performing the relevant
minimization, one finds however that the reduction in energy cost is
only about 10\%, with a considerable complication in the algorithm. In
this paper we have chosen to implement only the simplest (most local)
version of a coupled move-field update, corresponding to the situation
in Fig.~\ref{fig:plaquettes}.

A more general procedure, in which a heat-bath update on the combined
(particle move)+(field update) space provides a complete local
decorrelation between adjacent Monte Carlo configurations can easily
be derived as follows. We remind the reader that, in a heat-bath Monte
Carlo update, parameters are introduced to characterize a subspace of
the configuration space in the neighborhood of the starting
configuration, the dependence of the full Boltzmann weight of the
theory on these parameters is extracted (from the full Hamiltonian),
and then new values for these parameters are then chosen (independent
of the original configuration) on the basis of this Boltzmann
weight. In the situation considered here, the parameter space consists
of a single discrete particle move variable $m=0,1$ (with $m=0$
corresponding to no move, $m=1$ to a move along a specified link), and
four continuous plaquette shift variables $\alpha_{i},i=1,..4$.  For
the indicated environment of the central link $E_0$ in
Fig.~\ref{fig:plaquettes}, the relevant part of the Hamiltonian, as a
function of $m$ and the continuous plaquette update variables
$\alpha_{i}$, becomes
\begin{eqnarray}
\label{eq:Hfunc}
   {\cal H}(m,\alpha_{i}) &=& \frac{\hat{\beta}}{2} \Bigg\{
 (\hat{E}_0+\sum_{i=1}^{4}\alpha_{i}-mz)^{2} \label{eq:HMAlpha} \\
 &+&\sum_{l=1}^{3}(\hat{E}_{l}+\alpha_1)^{2} 
+\sum_{l=4}^{6}(\hat{E}_{l}+\alpha_2)^{2}\nonumber\\
&+&\sum_{l=7}^{9}(\hat{E}_{l}+\alpha_3)^{2}+\sum_{l=10}^{12}(\hat{E}_{l}+\alpha_4)^{2}\Bigg\}\nonumber
\end{eqnarray}
where we have introduced a variable (integer) valence $z$ to take care
of the case (needed in the simulations of Section 4) of multivalent
ions.  In order to implement a heat-bath procedure for this energy
function, we need to generate values for the quintet
$(m,\alpha_1,\alpha_2,\alpha_3,\alpha_4)$ distributed according to the
Boltzmann weight $e^{-{\cal H}(m,\alpha_{i})}$. Fortunately, a
complete analytic solution to this problem can easily be
derived. First, we note that the energy function in Eq.~\ref{eq:Hfunc}
can be reexpressed
\begin{eqnarray}
\label{eq:Hquad1}
 {\cal H}(m,\alpha_{i})
 &=&\frac{\hat{\beta}}{2}  \Bigg\{  (\hat{E}_{0}-mz)^{2}+\sum_{i,j}\alpha_{i}M_{ij}\alpha_{j} \\
 &&+ 2\sum_{i}\lambda_{i}\alpha_{i}+\sum_{l=1}^{12}\hat{E}_{l}^{2}
 \Bigg\} \nonumber
\end{eqnarray}
where 
\begin{eqnarray}
  \lambda_{i} &\equiv & P_{i}-mz \label{eq:lambda} \\
  P_1 &=& \hat{E}_0+\hat{E}_1+\hat{E}_2+\hat{E}_3 \label{eq:firstP}\\
  P_2 &=& \hat{E}_0+\hat{E}_4+\hat{E}_5+\hat{E}_6 \\
  &{\bf etc} \label{eq:etcP}&
\end{eqnarray}
i.e. the variables $P_{i}$, $i=1,2,3,4$ are just the plaquette fields obtained by summing
the electric link variables around each of the four plaquettes containing the central link
of Fig.~\ref{fig:plaquettes}, and the 4x4 matrix $M_{ij}$ 
\begin{equation}
 M_{ij} = 3\delta_{ij}+1
\end{equation}
is easily found to have inverse
\begin{equation}
  M^{-1}_{ij} = \frac{1}{3}(\delta_{ij}-\frac{1}{7})
\end{equation}

 Completing the square in Eq.~\ref{eq:Hquad1}, we find that ${\cal H}$ takes the form
\begin{eqnarray}
\label{eq:Hquad2}
 {\cal H} &=&
 \frac{\hat{\beta}}{2}\Bigg\{(\hat{E}_{0}-mz)^{2} + \sum_{ij} \xi_{i}M_{ij}\xi_{j} \\
&&-\sum_{ij} \lambda_{i}M^{-1}_{ij} \lambda_{j} +\sum_{l=1}^{12}\hat{E}_{l}^{2}\Bigg\}  \nonumber \\
  \xi_{i} &\equiv& \alpha_{i} + \sum_{j} M^{-1}_{ij}\lambda_{j}
\end{eqnarray}
The dependence of the local energy on the discrete move variable $m$
arises from the first and third terms in Eq.~\ref{eq:Hquad2} and the
corresponding Boltzmann weight determining the relative probability of
a particle move ($m=1$) versus no move ($m=0$) is therefore
\begin{equation}
    \exp{\left(\frac{\hat{\beta}}{2}\left(2\hat{E}_{0}z-\frac{3}{7}z^{2}-\frac{2}{7}z\sum_{i}P_{i}\right)m\right)}
\label{eq:BoltzmanWeightM}
\end{equation}
where we have used the fact that the move variable $m=0,1$ so that
$m^2=m$. A heat bath update of the variable $m$ is therefore trivial
to implement.

 The continuous $\alpha_{i}$ variables can be generated easily from
the Gaussian distribution of the $\xi_{i}$. The eigenvalues of
$M_{ij}$ are easily found (they are 7,3,3,3), as are the eigenvectors,
whence we find that the contribution of the second term in
Eq.~\ref{eq:Hquad2} to the Boltzmann weight can be rewritten
\begin{equation}
\label{eq:etaweight}
  \exp{\left(-\frac{\hat{\beta}}{2}\left(7\eta_{1}^{2}+3\eta_{2}^{2}+3\eta_{3}^{2}+3\eta_{4}^{2}\right)\right)}
\end{equation}
where
\begin{eqnarray}
\label{eq:eta1}
  \eta_{1} &=& \frac{1}{2}(\xi_{1}+\xi_{2}+\xi_{3}+\xi_{4}) \\
  \eta_{2} &=& \frac{1}{\sqrt{2}}(\xi_{1}-\xi_{2})  \\
  \eta_{3} &=& \frac{1}{\sqrt{2}}(\xi_{3}-\xi_{4})  \\
\label{eq:eta4}
  \eta_{4} &=& \frac{1}{2}(\xi_{1}+\xi_{2}-\xi_{3}-\xi_{4})
\end{eqnarray}
The heat-bath procedure for the plaquette shifts $\alpha_{i}$
therefore amounts to generating the independent Gaussian distributed
variables $\eta_{i}$ according to the weight (\ref{eq:etaweight}),
whence the $\alpha_{i}$ can be reconstructed via Equations
\ref{eq:eta1}-\ref{eq:eta4} and
\begin{eqnarray}
   \alpha_{i} &=& \xi_{i} - \sum_{j} M^{-1}_{ij}\lambda_{j}  \\
   &=& \xi_{i} -\frac{1}{3}\lambda_{i} +
   \frac{1}{21}\sum_{j}\lambda_{j} \label{eq:toCalcAlpha}
\end{eqnarray}

 To summarize, the algorithm for a coupled particle/field heat-bath  update is implemented 
as follows:
\begin{enumerate}
\item Calculate the plaquette sums $P_{i},i=1,2,3,4$ 
(Equations \ref{eq:firstP}-\ref{eq:etcP}) for the four plaquettes 
interfacing the link along which we desire to move the particle.
\item Choose the move variable $m=0,1$ with weight given 
by Eq.~\ref{eq:BoltzmanWeightM}.
\item Generate independent Gaussian variables $\eta_{i},i=1,2,3,4$ 
according to Eq.~\ref{eq:etaweight}.
\item Solve Equations \ref{eq:eta1}-\ref{eq:eta4} for the $\xi_{i},i=1,2,3,4$.
\item Compute $\lambda_{i}$ from Eq.~\ref{eq:lambda} and use Eq.~\ref{eq:toCalcAlpha} 
to obtain the desired plaquette
shifts $\alpha_{i},i=1,2,3,4$, which are then used to update the 
electric fields $\hat{E}_{l},l=0,..12$
as indicated in Eq.~\ref{eq:HMAlpha} (see Fig.~\ref{fig:plaquettes}).  
\item If $m=1$ then move the particle across the considered link
while updating the electric field on the link according 
to Eq.~\ref{eq:particlemove}.
\end{enumerate}

Recently the problem of low acceptance rates for particle moves was
noted by Maggs et al.\ in Ref.~\onlinecite{maggs:acc_rate}.  They present an
alternative solution to the problem where each charge, instead of
residing on a single lattice site, is broken into pieces and resides
on the lattice sites in an $n$x$n$x$n$ cube.  In order to move a
particle, all of the pieces of the particle must be moved in unison.
They have shown that the inverse temperature that they are able to
simulate efficiently grows like $n^3$ using this method.  The
advantage of this method is that it is effective at increasing the
acceptance rate and that the size of the cube can be chosen to give
the desired acceptance rate. The disadvantage of this method is that
the charges are spread out so that, for systems that are sensitive to
the spatial location of the charges, the lattice must be made finer by
a factor of $n$ in every direction to obtain the same charge locality
as the lattice with unbroken particles.  Using the methods discussed
in this work, the charges remain on a single lattice site so there are
no difficulties arising from the breakup of the ions onto different
lattice sites.

\section{Applications: Strongly Fluctuating Fields between Charged Plates/Membranes}

To test these algorithms on a strongly charged system where
correlation effects play a major role, we have considered a system of
charged conducting plates with ions between the plates.  While the
system is electrically neutral, there is an osmotic pressure between
the plates that depends on the electrostatic interaction between the
particles and on the correlations between the particles.  This system
has been extensively studied both
theoretically\cite{podgornik88,stevens90} and
numerically\cite{guldbrand84,moreira02} and is known to be a strongly
fluctuating system within the parameter ranges in which we are
interested.  A convenient criterion\cite{moreira02} for a strongly
fluctuating system is that the Bjerrum length, $\ell_B$, times the
square of the ion valence, $z$, be smaller then the Gouy-Chapman
length, $\mu$. In the systems we are considering $z^2\ell_B / \mu$ is
as large as $33$.  We have considered both divalent and univalent
ions, as in previous work it was seen that there is a repulsive
pressure in the univalent case and an \textit{attractive} pressure in
the divalent case\cite{guldbrand84,podgornik88,moreira02}.  It is
shown in Ref.~\onlinecite{podgornik88} that the Poisson-Boltzmann
calculations of the osmotic pressure in the divalent case break down
and cannot even predict the sign of the osmotic pressure.

Our basic system consists of a 50x50x50 lattice with periodic boundary
conditions in all three dimensions.  Positive charges are free to move
on two fixed plates separated in the $x$ direction which extend the
entire extent of the lattice in the $y$ and $z$ directions, while the
region between the plates contains mobile counterions ensuring overall
neutrality.  The periodicity in the $x$ direction is not critical, as
quantities observed are insensitive to field fluctuations far outside
the plates.  We choose a lattice spacing of 1 \AA, so that we can
study plate separations in the range of interest.  Using the
dielectric constant of water ($\epsilon = 80.0$) and room temperature
($T = 300 K$) gives a dimensionless inverse temperature $\hat{\beta} =
87.1$, too large to effectively simulate with simple particle moves
that do not adjust the electric field on neighboring plaquettes.  We
placed 34 positive univalent charges on each of the plates to give a
surface-charge density of $0.2176\, C\, m^{-2}$, approximately that
used in References~\onlinecite{guldbrand84,podgornik88}.  These charges are
allowed to move during the simulation, but are not allowed to leave
the plate.  To make the system electrically neutral, 68 negatively
charged ions are placed between the plates in the univalent case, and
34 negatively charged ions in the divalent case.  Two ions are
forbidden from being on the same lattice site.  The charges on the
plates are initially randomly distributed on the plates, and the ions
between the plates are initially distributed with half of the ions on
the closest allowed plane to the right plate, and half on the closest
allowed plane to the left plate.  All runs are composed of 5,000 Monte
Carlo equilibration steps followed by 20,000 measurement steps.  Each
Monte Carlo step is composed of a coupled Metropolis update of the
electric field around each plaquette, (200 $\times$ Number of charges
on the plates) attempted moves of a particle on the plate chosen at
random, and (20000 $\times$ Number of charges in solution) attempted
moves of a particle in solution. As pointed out by Maggs et
al.\cite{maggs:dielectric}, a global update of the electric field is
also included to ensure rigorous ergodicity.

To investigate the importance of the mobility of the charges on the
plates we have also performed a set of simulations with the positive
charges on plates fixed at a random initial distribution.  There were
no qualitative differences between the results of these simulations
and the results of the simulations with mobile ions on the plates that
we present here.

To investigate the errors due to lattice effects, we have also studied
an asymmetric lattice where the lattice spacing is a factor of two
smaller in the dimension separating the plates.  This is a 100x50x50
asymmetric lattice with a lattice spacing of 0.5 \AA\ in the $x$
direction so that the total volume of the system remains constant.
Two ions are again forbidden from being on the same lattice site.  In
the asymmetric case this corresponds to a different hard sphere
interaction between the ions than in the symmetric lattice, but these
differences are in practice unimportant, as the ions are so sparsely
distributed that collision between ions are rare.

The ions between the plates will naturally accumulate on the planes of
lattice sites close to the plates.  As the electric potential changes
rapidly in this region, the results of our simulation will depend on
the details of the discretization in this region. As the discreteness
of the lattice has the largest effect in the region close to the
plates, we have chosen to forbid the ions from coming within 1 \AA\ of
the plates.  This will soften slightly the potential seen by the ions.
On the symmetric lattice, we do not allow ions on the planes of
lattice sites closest to the plates.  On the asymmetric lattice, ions
are not allowed on the 2 planes of lattice sites closest to the
plates.

We are primarily interested in observing the osmotic pressure between
the plates as we change the separation, both for the univalent and
divalent ions in solution.  As derived in
Ref.~\onlinecite{guldbrand84}, the osmotic pressure can be calculated
using the expression
\begin{equation}
p_{\mbox{osm}}= kT\, C(0) + F^{AB}_x/area,
\label{eq:pressure}
\end{equation}
where $C(0)$ is the ion concentration at the mid-plane and $F^{AB}_x$
is the average electrostatic force between the left half of the system and the right
half of the system.  In the continuum, this force could be written as
\begin{equation}
\label{eq:fxequation}
F^{AB}_x = \frac{1}{\epsilon} \sum_m^A \sum_n^B q_n q_m \Delta x_{mn} / r^3_{mn},
\end{equation}
where $A$ is the set of all charges to the left of the mid-plane, $B$
is the set of all charges to the right of the mid-plane, $\Delta x_{mn}$
is the separation between the charges in the $x$ direction, and $r_{mn}$
is the distance between the charges.  In order to take into account
lattice effects and correctly treat the periodic boundary conditions,
we have calculated the force using the lattice coulomb force. 
This is done simply by replacing the continuum quantity $\Delta x_{mn}/r^{3}_{mn}$
in Eq.~\ref{eq:fxequation} by the corresponding lattice expression
\begin{equation}
   \frac{4\pi}{L^{3}}\sum_{\vec{k}\ne 0}\frac{i\sin{(2\pi k_{x}/L)}e^{2\pi i\vec{k}\cdot\vec{r}_{mn}}}
{4\sum_{i=1}^{3}\sin^{2}{(\pi k_{i}/L)}}
\end{equation}
where $k_{i}=0,1,2,....L-1$.

  Although the osmotic pressure could be
calculated by observing the change in free energy as the separation of the
plates is changed, work in Ref.~\onlinecite{guldbrand84} has shown that the
pressure calculated using Eq.~\ref{eq:pressure} has fewer fluctuations
than the pressure calculated using the free energy difference.

%
\begin{figure}
\centerline{\includegraphics[width=3.5in]{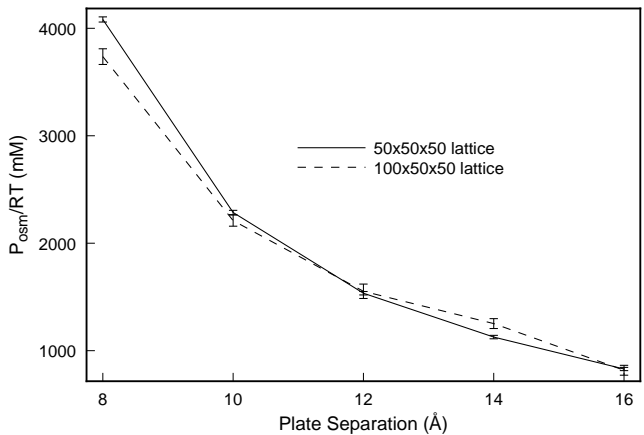}}
\caption{
Osmotic pressure from simulations of univalent ions at a range of
plate separations.}
\label{fig:univalent-pressure}
\end{figure}   
\begin{figure}
\centerline{\includegraphics[width=3.5in]{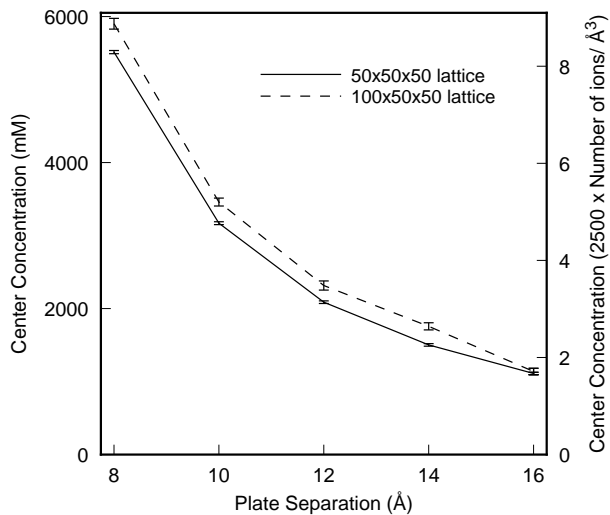}}
\caption{
Concentration of charges at the plane between the charges from
simulations of univalent ions.  Left axis has the same units as the
pressure results.  Right axis gives the number of charges on the
center plane (50x50x50 lattice) or center two planes (100x50x50
lattice).  This is one of the terms contributing to the osmotic
pressure.}
\label{fig:univalent-concentration}
\end{figure}
\begin{figure}
\centerline{\includegraphics[width=3.5in]{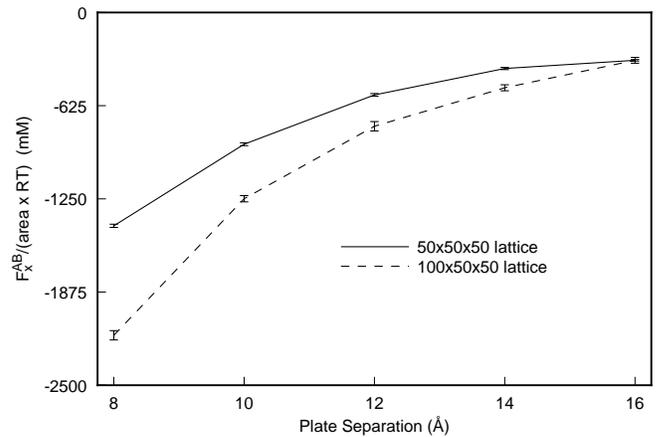}}
\caption{
Electrostatic force between the halves of the system for univalent
ions. This is the second term contributing to the osmotic pressure.}
\label{fig:univalent-force-term}
\end{figure}   
\begin{figure}
\centerline{\includegraphics[width=3.5in]{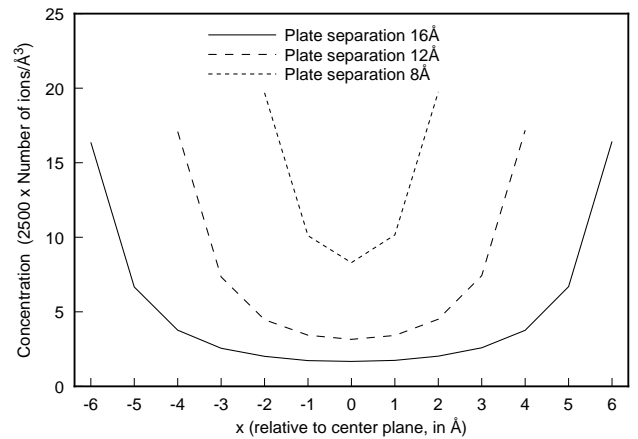}}
\caption{Concentration profile of the univalent ions in solution for a range 
of plate separations.  For each $x$, the average number of ions in the
system a distance $x$ from the center plate is shown. The ion concentration is 0 outside of the range shown.}
\label{fig:conc_prof_uni}
\end{figure}   

The pressure for univalent ions is shown in
Fig.~\ref{fig:univalent-pressure} for a range of plate separations.  Results
are shown from both the 50x50x50 lattice and the 100x50x50 lattice.
The ions are moved using the coupled Metropolis update method
described in Sec.~\ref{sec:updateproc}, as the heat bath method is
difficult to adapt to the asymmetric lattice.
Figure~\ref{fig:univalent-concentration} shows the first term of
Eq.~\ref{eq:pressure}, the concentration of ions on the plane between
the plates, for the same simulations.  The left axis of this plot
gives the average number of ions in a 1 \AA\ by 50 \AA\ by 50 \AA\
rectangular box centered between the plates.
Figure~\ref{fig:univalent-force-term} shows the second term of
Eq.~\ref{eq:pressure}.  The differences in results from the two
lattice sizes are modest, showing that the errors due to
lattice discretization are small. Figure~\ref{fig:conc_prof_uni} shows the
concentration profiles of the ions in solution from the simulations on
the 50x50x50 lattice.
The ions are attracted to the plates, but a small density of ions
remains in the center of the gap between the planes.


\begin{figure}
\centerline{\includegraphics[width=3.5in]{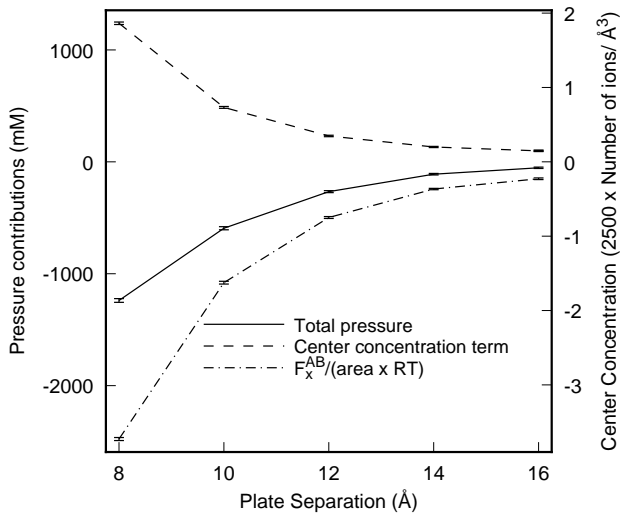}}
\caption{ Results from simulations on divalent ions at a range of
plate separations on a 50x50x50 lattice.  Shown with a solid line is
the total osmotic pressure divided by $RT$. Shown with a dashed line
and a dotted-dashed line are the two terms that
contribute to it, the ion concentration on the center plane and the
electrostatic force between the two halves of the system.  The axis on
the left gives the values in units of micromolars; the axis on the right
gives the concentration on the center plane in terms of the number of ions in our system on the center plane.}
\label{fig:divalent}
\end{figure}   

\begin{figure}
\centerline{\includegraphics[width=3.5in]{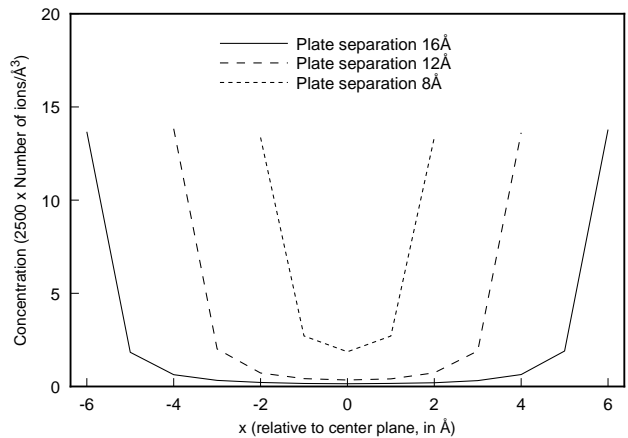}}
\caption{Concentration profile of the divalent ions in solution for a range of plate separations.  For each $x$, the average number of ions in the
system a distance $x$ from the center plate is shown.  The ion concentration is 0 outside of the range shown.}
\label{fig:conc_prof_dv}
\end{figure}   

For the divalent ions we only consider the 50x50x50 lattice.  Here we
use the heat-bath method for moving the particles and updating the
electric fields.  There are 10,000 equilibration steps and 200,000
measurement steps, and other parameters are the same as the univalent
case.  The solid line in Fig.~\ref{fig:divalent} shows the pressure in
the divalent case, while dashed line and dashed-dotted line show the
first and second terms of Eq.~\ref{eq:pressure}. Qualitatively, our
result agree with previous theoretical and Monte Carlo work, although
a direct quantitative comparison is difficult because of the
differences in how the interaction with the plate is treated.  The
concentration profiles of the ions in solution for these simulations
are shown in Fig.~\ref{fig:conc_prof_dv}.  Note that the ions are much
more tightly bound to the plates in the divalent case as compared with
the univalent case.

\begin{figure}
\centerline{\includegraphics[width=3.5in]{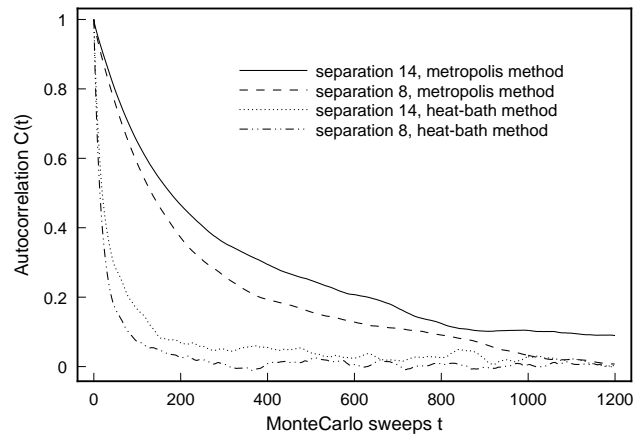}}
\caption{Comparison of pressure autocorrelation functions using the coupled
Metropolis move update, and the heat-bath particle move update. 
These autocorrelation functions are all from simulations with divalent ions.}
\label{fig:autocorr}
\end{figure}   

The simulations discussed here would be difficult or impossible to
perform with the simple particle Metropolis move, which does not adjust
the field on neighboring plaquettes.  By using the coupled Metropolis
updating move described in this work we are able to increase the
acceptance rate for particle moves in the univalent case to 0.13 from
$10^{-4}$, and in the divalent case to 0.0056 from less than $10^{-6}$.
The anisotropy of the 100x50x50 lattice affects the acceptance rates.
In this case the univalent acceptance rates using the coupled
Metropolis move increases to 0.48 in the $x$ direction, but the
acceptance rate for particle moves perpendicular to the $x$ direction
drops to 0.0061.  This asymmetry occurs because the coupled update
procedure is no longer able to effectively spread the change in
electric field when a particle is moved in the $y$ or $z$ directions.

When using the heat-bath approach to moving the particles, the
acceptance rate (0.10 for univalent ions and 0.0045 for divalent ions) is
lower than the coupled Metropolis move acceptance rate, but still has
a much greater acceptance rate than the simple Metropolis particle
move that does not adjust the electric field on neighboring
plaquettes.  The advantage of the
heat-bath approach is that it better decorrelates the system so that the
observables have a shorter autocorrelation time.  The autocorrelation
function of a observable $A_t$ is given by
\begin{equation}
C(t) = \frac{1}{N-t} \sum_{j=1}^{N-t} (A_j - \bar{A})(A_{j+t} -\bar{A}),
\end{equation}
where $\bar{A}$ is the average of $A_t$.  The autocorrelation function
of the pressure is shown in Fig.~\ref{fig:autocorr} for plate
separations of 8 and 14 with divalent ions using both the coupled
Metropolis particle move update and the heat-bath particle move
update.  Although the autocorrelation time (obtained by integrating
the autocorrelation function) increases with the larger plate
separation, the autocorrelation time with the heat-bath update is
consistently smaller than the autocorrelation time for the coupled
Metropolis update.  Across all plate separations, the autocorrelation
times from heat-bath updates were five to ten times smaller than the
autocorrelation times from coupled Metropolis updates.  This more than
compensates for the additional computational cost per sweep
(approximately twice that of the coupled Metropolis update) of the
heat-bath update.

\section{Conclusions}

The development of efficient local algorithms for Monte Carlo
simulation of Coulomb systems with non-uniform dielectric constants is
crucial for the study of the larger and more physically realistic
biophysical systems of interest to researchers\cite{Zacharias,
  Bashford, Spassov}.  The technique
of Maggs et al.\ shows great promise in fulfilling these goals, but it
must be shown to be efficient and accurate in physically interesting
parameter ranges.  Studying a system of parallel plates screened by
ion with a large dimensionless inverse temperature, we see that the
simplest method of moving particles, where the electric field is only
modified on the link traversed by the particle, gives unusably small
acceptance rates.  By updating the electric field on plaquettes
neighboring the traversed link, we can increase the acceptance rates to
a usable level.  Using a heat bath approach reduces the
autocorrelation time of the simulation.

\section{Acknowledgments}

The work of A.~Duncan was
supported in part by NSF grant PHY0244599. The work of R.D.~Sedgewick
and R.D.~Coalson was supported by NSF grant CHE0092285.


\begin{thebibliography}{99}
\bibitem{maggs:prl} A.C.~Maggs and V.~Rossetto, Phys.\ Rev.\ Lett.\ \textbf{88}, 196402 (2002)
\bibitem{bluegene} F.~Allen et al., IBM Systems Journal \textbf{40},
310 (2001)
\bibitem{maggs:acc_rate} L.~Levrel, F.~Alet, J.~Rottler, and
A.C.~Maggs, Statphys22 Proceedings, to be published in PRAMANA 
[also available at  \texttt{cond-mat/0409350}]
\bibitem{podgornik88} R.~Podgornik and B.~\v{Z}ek\v{s}, J.\ Chem.\
Soc., Faraday Trans. 2 \textbf{84}, 611 (1988)
\bibitem{stevens90} M.J.\ Stevens and M.O.\ Robbins, 
Europhys.\ Lett.\ \textbf{12}, 81 (1990)
\bibitem{guldbrand84} L.~Guldbrand, B.~J\"onsson, H.~Wennerstr\"om,
and P.~Linse, J.\ Chem.\ Phys.\ \textbf{80}, 2221 (1984)
\bibitem{moreira02} A.G.~Moreira and R.R.~Netz, Eur.\ Phys.\ J.\ E
  \textbf{8}, 33 (2002)
\bibitem{Leach} For a summary of these techniques see: A.~Leach, 
``Molecular Modeling: Principles and Applications,'' Prentice Hall,
2001
\bibitem{Zacharias} M.\ Zacharias, B.A.\ Luty, M.E.\ Davis and J.A.\ McCammon, 
J.\ Mol.\ Biol.\ \textbf{238}, 455 (1994)
\bibitem{Bashford} D.\ Bashford and M.\ Karplus, Biochemistry \textbf{29}, 10219 (1990)
\bibitem{Spassov} V.\ Spassov and D.\ Bashford, Protein Sci.\ \textbf{7}, 2012 (1998)
\bibitem{Edwards59} S.F.\ Edwards, Philos.\ Mag.\ \textbf{4}, 1171 (1959);
\bibitem{Edwards62} S.F.\ Edwards and A.\ Lenard, 
J.\ Math.\ Phys.\ \textbf{3}, 778 (1962)
\bibitem{tonyrob} R.~Coalson and A.~Duncan, J.\ Chem.\ Phys.\
 \textbf{97}, 5653 (1992)
\bibitem{LQCD} I.~Montray and G.~M\"unster, ``Quantum Fields on a
lattice,'' Cambridge Monographs on Mathematical Physics, 1997
\bibitem{maggs:dielectric} A.C.~Maggs, J.~Chem.~Phys. \textbf{120}, 3108 (2004)
\bibitem{NR} W.H.~Press, B.P.~Flannery, S.A.~Teukolsky, and
W.T.~Vetterling, ``Numerical Recipes in C: The Art of Scientific
Computing,'' Cambridge University Press, 1992
\end{thebibliography}
\end{document}